\documentclass[zpreprint,zbstdefault]{zeus_paper}
\usepackage[english]{babel}
\newcommand{\Zdetdesc}{%
A detailed description of the ZEUS detector can be found 
elsewhere~\cite{zeus:1993:bluebook}. A brief outline of the 
components that are most relevant for this analysis is given
below.\xspace}
\newcommand{\Zcaldesc}{%
The high-resolution uranium--scintillator calorimeter (CAL)~\citeCAL consists 
of three parts: the forward (FCAL), the barrel (BCAL) and the rear (RCAL)
calorimeters. Each part is subdivided transversely into towers and
longitudinally into one electromagnetic section (EMC) and either one (in RCAL)
or two (in BCAL and FCAL) hadronic sections (HAC). The smallest subdivision of
the calorimeter is called a cell.  The CAL energy resolutions, as measured under
test-beam conditions, are $\sigma(E)/E=0.18/\sqrt{E}$ for electrons and
$\sigma(E)/E=0.35/\sqrt{E}$ for hadrons ($E$ in $\Gev$).}
\chardef\usc=95
\chardef\til=126
\catcode`\@=11 % @ signs are now treated as letters
\DeclareRobustCommand\xdotspace{\futurelet\@let@token\@xdotspace}
\def\@xdotspace{%
  \ifx\@let@token.\else
  \ifx\@let@token\bgroup.\else
  \ifx\@let@token\egroup.\else
  \ifx\@let@token\/.\else
  \ifx\@let@token\ .\else
  \ifx\@let@token~.\else
  \ifx\@let@token!.\else
  \ifx\@let@token,.\else
  \ifx\@let@token:.\else
  \ifx\@let@token;.\else
  \ifx\@let@token?.\else
  \ifx\@let@token/.\else
  \ifx\@let@token'.\else
  \ifx\@let@token).\else
  \ifx\@let@token-.\else
  \ifx\@let@token\@xobeysp.\else
  \ifx\@let@token\space.\else
  \ifx\@let@token\@sptoken.\else
   .\space
   \fi\fi\fi\fi\fi\fi\fi\fi\fi\fi\fi\fi\fi\fi\fi\fi\fi\fi}
\catcode`\@=12 % @ signs are no longer letters

\newcommand{\stru}[2]{%
   \relax\ifmmode\hbox{\vrule height#1 depth#2 width0pt}%
   \else\vrule height#1 depth#2 width0pt\fi}
\newcommand{\Ronum}[1]{\uppercase\expandafter{\romannumeral#1}}
\newcommand{\ronum}[1]{\expandafter{\romannumeral#1}}
\DeclareRobustCommand{\LaTeXZ}{%
  \LaTeX\kern-.05em4\kern-.1em
  {\raisebox{-0.2ex}{$\scriptstyle\text{ZEUS}$}}\xspace}
\DeclareMathAlphabet{\mathbf}{OT1}{cmr}{bx}{sl}
\newcommand{\eVdist}{\kern-0.06667em}

\newcommand{\Gev}{{\text{Ge}\eVdist\text{V\/}}}

\newcommand{\gev}{{\,\text{Ge}\eVdist\text{V\/}}}

\newcommand{\Tesla}{\,\text{T}}

\newcommand{\slashfrac}[2]{%
  \raisebox{0.5ex}{\ensuremath #1}\kern-0.12em/\kern-0.08em
  \raisebox{-.8ex}{\ensuremath #2}}
\newcommand{\sqr}[3]{%
    {\vcenter{\hrule height.#3ex\hbox{\vrule width.#2ex height#1ex
     \kern#1ex\vrule width.#3ex}\hrule height.#2ex}}}

\catcode`\@=11 % @ signs are now treated as letters
\newcommand{\parenbar}{\mathpalette\p@renb@r}
\def\p@renb@r#1#2{\vbox{%
  \ifx#1\scriptscriptstyle \dimen@.7em\dimen@ii.2em\else
  \ifx#1\scriptstyle \dimen@.8em\dimen@ii.25em\else
  \dimen@1em\dimen@ii.4em\fi\fi \offinterlineskip
  \ialign{\hfill##\hfill\cr
    \vbox{\hrule width\dimen@ii}\cr
    \noalign{\vskip-.3ex}%
    \hbox to\dimen@{$\mathchar300\hfil\mathchar301$}\cr
    \noalign{\vskip-.3ex}%
    $#1#2$\cr}}}
\catcode`\@=12 % @ signs are no longer letters

\newcommand{\IP}{{\rm I$\kern-0.01667em$P}\xspace}

\mathchardef\qsm=63
\mathchardef\pls=43
\mathchardef\mns=512
\mathchardef\plm=518
\mathchardef\eql=61
\mathchardef\smallleft=300
\mathchardef\smallright=301
\mathchardef\les=316
\mathchardef\gre=318
\mathchardef\leq=532
\mathchardef\grq=533
\catcode`\@=11 % @ signs are now treated as letters
\newcounter{pict@width}
\newcounter{pict@height}
\newlength{\pict@scale}
\newcommand{\psfigadd}[4]{%
\setcounter{pict@width}{1*\ratio{#2+\pict@scale/2}{\pict@scale}}
\setcounter{pict@height}{1*\ratio{#3+\pict@scale/2}{\pict@scale}}
\setlength{\unitlength}{\pict@scale}
\hbox to #2{\hspace{-\fill}\begin{picture}(\thepict@width,\thepict@height)
\put(0,0){\psfig{figure=#1,width=#2,height=#3,clip=}}
\SetScale{0.283466457}
\SetWidth{1.763889}
{#4}
\end{picture}}
}
\newcounter{pict@widthfst}
\newcounter{pict@widthscd}
\newcounter{pict@widthtot}
\newcommand{\psfigaddtwo}[7]{%
\setcounter{pict@widthfst}{1*\ratio{#2+\pict@scale/2}{\pict@scale}}
\setcounter{pict@widthscd}{1*\ratio{#2+#4+\pict@scale/2}{\pict@scale}}
\setcounter{pict@widthtot}{1*\ratio{#2+#4+#6+\pict@scale/2}{\pict@scale}}
\setcounter{pict@height}{1*\ratio{#3+\pict@scale/2}{\pict@scale}}
\setlength{\unitlength}{\pict@scale}
\hbox{\hspace{-\fill}\begin{picture}(\thepict@widthtot,\thepict@height)
\put(0,0){\psfig{figure=#1,width=#2,height=#3,clip=}}
\put(\thepict@widthscd,0){\psfig{figure=#5,width=#6,height=#3,clip=}}
\SetScale{0.283466457}
\SetWidth{1.763889}
{#7}
\end{picture}}
}
\newcommand{\psfigror}[4]{%
\setcounter{pict@width}{1*\ratio{#2+\pict@scale/2}{\pict@scale}}
\setcounter{pict@height}{1*\ratio{#3+\pict@scale/2}{\pict@scale}}
\setlength{\unitlength}{\pict@scale}
\hbox{\begin{picture}(\thepict@width,\thepict@height)
\put(0,\thepict@height){\psfig{figure=#1,width=#3,height=#2,clip=,angle=270}}
\SetScale{0.283466457}
\SetWidth{1.763889}
{#4}
\end{picture}}
}
\newcommand{\psfigrol}[4]{%
\setcounter{pict@width}{1*\ratio{#2+\pict@scale/2}{\pict@scale}}
\setcounter{pict@height}{1*\ratio{#3+\pict@scale/2}{\pict@scale}}
\setlength{\unitlength}{\pict@scale}
\hbox{\begin{picture}(\thepict@width,\thepict@height)
\put(0,0){\psfig{figure=#1,width=#3,height=#2,clip=,angle=90}}
\SetScale{0.283466457}
\SetWidth{1.763889}
{#4}
\end{picture}}
}
\catcode`\@=12 % @ signs are no longer letters
\newlength\listtextwidth

\catcode`\@=11 % @ signs are now treated as letters
\newlength{\@tabfninsert}
\newlength{\@tabfnwidth}
\newcommand{\tabfootnote}[2]{%
  \setlength{\@tabfninsert}{0.8em}
  \setlength{\@tabfnwidth}{\textwidth}
  \addtolength{\@tabfnwidth}{-\@tabfninsert}
  \addtolength{\@tabfnwidth}{-0.4em}
  \noindent\makebox[\@tabfninsert][r]{\footnotesize$^{#1}$\hfil}\hfill%
  \parbox[t]{\@tabfnwidth}{\footnotesize #2\hfill}}
\catcode`\@=12 % @ signs are no longer letters

\def\citeCTD{{\cite{%
nim:a279:290,*npps:b32:181,*nim:a338:254%
}}\xspace}
\def\citeMVD{{\cite{%
nim:a581:656%
}}\xspace}
\def\citeCAL{{\cite{%
nim:a309:77,*nim:a309:101,*nim:a321:356,*nim:a336:23%
}}\xspace}

\begin{document}
\prepnum{{DESY--09--036}}
\date{March 2009}
%------------------------------------------------------------------------------
%       Title sheet
%------------------------------------------------------------------------------
\title{Exclusive photoproduction of {\bf $\Upsilon$} mesons at HERA}                                                       
                    
\author{ZEUS Collaboration}
\draftversion{8.0}
\date{27\ January 2009}

\abstract{
The exclusive photoproduction reaction $\gamma\:p \rightarrow
\Upsilon \:p$ has been studied with the ZEUS experiment in $ep$ collisions
at HERA using an integrated luminosity of 468~pb$^{-1}$. The
measurement covers the kinematic range $60<W<220$~GeV and $Q^2<1$
GeV$^2$, where $W$ is the photon-proton centre-of-mass energy and
$Q^2$ is the photon virtuality.  These results, which represent the
analysis of the full ZEUS data sample for this channel, are compared
to predictions based on perturbative QCD.  }

\makezeustitle

\newpage

\def\3{\ss}
\pagenumbering{Roman}
%===================================================================                               
%                                                                                                  
%  MEMBER NAME  AUTH156a (ZEUS)     M  TEX                                                         
%                                                                                                  
%  JH.: transformed to a format, which is suited as input for                                      
%       CONVERT, which automatically creates author-indices                                        
%                                                                                                  
%  Don't remove lines starting with a percent sign %,                                              
%  CONVERT may need them urgently !                                                                
%                                                                                                  
%=====================================================================                             
\begin{center}                                                                                     
{                      \Large  The ZEUS Collaboration              }                               
\end{center}                                                                                       
  S.~Chekanov,                                                                                     
  M.~Derrick,                                                                                      
  S.~Magill,                                                                                       
  B.~Musgrave,                                                                                     
  D.~Nicholass$^{   1}$,                                                                           
  \mbox{J.~Repond},                                                                                
  R.~Yoshida\\                                                                                     
 {\it Argonne National Laboratory, Argonne, Illinois 60439-4815, USA}~$^{n}$                       
\par \filbreak                                                                                     
  M.C.K.~Mattingly \\                                                                              
 {\it Andrews University, Berrien Springs, Michigan 49104-0380, USA}                               
\par \filbreak                                                                                     
  P.~Antonioli,                                                                                    
  G.~Bari,                                                                                         
  L.~Bellagamba,                                                                                   
  D.~Boscherini,                                                                                   
  G.~Bruni,                                                                                        
  F.~Cindolo,                                                                                      
  M.~Corradi,                                                                                      
\mbox{G.~Iacobucci},                                                                               
  A.~Margotti,                                                                                     
  R.~Nania,                                                                                        
  A.~Polini\\                                                                                      
  {\it INFN Bologna, Bologna, Italy}~$^{e}$
% ...
\newpage                                                                                           
\enlargethispage{5cm}                                                                              
$^{\    1}$ also affiliated with University College London,                                        
United Kingdom\\                                                                                   
$^{\    2}$ now at University of Salerno, Italy \\                                                 
% ...
$^{\    5}$ supported by the research grant No. 1 P03B 04529 (2005-2008) \\                        
% ...
%                                                                                                  
% \par         % if index listing & table fit to 1 page, put gap here                              
\newpage   % alternatively: go to newpage, if page is too small                                    
                                                           %                                       
% \institute_references_start    % do not touch or move this line !                                
                                                           %                                       
\begin{tabular}[h]{rp{14cm}}                                                                       
$^{a}$ &  supported by the Natural Sciences and Engineering Research Council of Canada (NSERC) \\  
$^{b}$ &  supported by the German Federal Ministry for Education and Research (BMBF), under        
          contract Nos. 05 HZ6PDA, 05 HZ6GUA, 05 HZ6VFA and 05 HZ4KHA\\                            
$^{c}$ &  supported in part by the MINERVA Gesellschaft f\"ur Forschung GmbH, the Israel Science   
          Foundation (grant No. 293/02-11.2) and the US-Israel Binational Science Foundation \\    
$^{d}$ &  supported by the Israel Science Foundation\\                                             
$^{e}$ &  supported by the Italian National Institute for Nuclear Physics (INFN) \\                
$^{f}$ &  supported by the Japanese Ministry of Education, Culture, Sports, Science and Technology 
          (MEXT) and its grants for Scientific Research\\                                          
$^{g}$ &  supported by the Korean Ministry of Education and Korea Science and Engineering          
          Foundation\\                                                                             
$^{h}$ &  supported by the Netherlands Foundation for Research on Matter (FOM)\\                   
$^{i}$ &  supported by the Polish State Committee for Scientific Research, project No.             
          DESY/256/2006 - 154/DES/2006/03\\                                                        
$^{j}$ &  partially supported by the German Federal Ministry for Education and Research (BMBF)\\   
$^{k}$ &  supported by RF Presidential grant N 1456.2008.2 for the leading                         
          scientific schools and by the Russian Ministry of Education and Science through its      
          grant for Scientific Research on High Energy Physics\\                                   
$^{l}$ &  supported by the Spanish Ministry of Education and Science through funds provided by     
          CICYT\\                                                                                  
$^{m}$ &  supported by the Science and Technology Facilities Council, UK\\                         
$^{n}$ &  supported by the US Department of Energy\\                                               
$^{o}$ &  supported by the US National Science Foundation. Any opinion,                            
findings and conclusions or recommendations expressed in this material                             
are those of the authors and do not necessarily reflect the views of the                           
National Science Foundation.\\                                                                     
$^{p}$ &  supported by the Polish Ministry of Science and Higher Education                         
as a scientific project (2006-2008)\\                                                              
$^{q}$ &  supported by FNRS and its associated funds (IISN and FRIA) and by an Inter-University    
          Attraction Poles Programme subsidised by the Belgian Federal Science Policy Office\\     
$^{r}$ &  supported by an FRGS grant from the Malaysian government\\                               
\end{tabular}                                                                                      
                                                           %                                       
% \institute_references_end     % do not touch or move this line !             

%------------------------------------------------------------------------------
%       Text
%------------------------------------------------------------------------------
\pagenumbering{arabic}
\pagestyle{plain}
% ----------------------------------------------------------------------------
%       Introduction
% ----------------------------------------------------------------------------
\section{Introduction}

Exclusive photoproduction of heavy vector mesons, $J/\psi$ and
$\Upsilon$, has previously been studied at
HERA~\cite{epj:c24:345,pl:b437:432,pl:b483:23,H1jpsi:2006}.  The
process $\gamma\:p \rightarrow V\:p$, with $V=J/\psi, \Upsilon$ can be
described by perturbative QCD (pQCD), since the relatively high masses
of the charm and bottom quarks provide sufficiently hard scales.  The
process is assumed to exhibit threefold factorisation
~\cite{Frankfurt:1998yf,Martin:2007sb,Ivanov:2004nk,rss}: the photon
fluctuates into a $q\bar{q}$ pair; the pair interacts with the proton;
and finally the heavy meson is formed in the final state.  At leading
order (LO), the interaction of the $q\bar{q}$ pair with the proton
proceeds via the exchange of two gluons in a colour-singlet
state. Thus the cross section is proportional to the square of the
gluon density in the proton. The rise of the gluon density with
decreasing fractional momentum, $x$, leads to the prediction of a
cross section rapidly rising as a function of the photon-proton
centre-of-mass energy, $W$, where the relevant $x$ region accessible
in heavy-quark production at HERA is $10^{-4}<x<10^{-2}$. The rise of
the cross section, $\sigma$, with $W$ can be expressed as $\sigma
\propto W^\delta$, where a value of $\delta\approx$ 0.7$-$0.8 has been
measured for the $J/\psi$~\cite{epj:c24:345,H1jpsi:2006}.  A value of
1.7 is predicted for $\Upsilon$(1S) production in
LO~\cite{Frankfurt:1998yf}.

Prior to this analysis, ZEUS and H1 measured the $\Upsilon$
photoproduction cross section for one value of $W$. The increased
statistics of the data used in this study allows the investigation of
the dependence of the production cross section on the energy $W$.  The
data cover the kinematic range $60<W<220$~GeV and the results are
obtained using the $\mu^+ \mu^-$ decay channel. In this measurement
the three upsilon states $\Upsilon$(1S), $\Upsilon$(2S) and
$\Upsilon$ (3S) (denoted $\Upsilon_i$, $i$=1,2,3, respectively) are
not resolved. Hence the sum of the cross sections multiplied by the
corresponding decay branching ratios to muons, $\sum_i \sigma^{
\gamma p \rightarrow \Upsilon_i p} \cdot \mathcal{B}_i $, was measured as a function of $W$
for two intervals: $60<W<130$~GeV and $130<W<220$~GeV.  The sample
under study represents more than a ten-fold increase in integrated
luminosity compared to the previous ZEUS publication
\cite{pl:b437:432}.

\section{Experimental set-up}
In 1998--2007 (1996--1997), HERA provided electron\footnote{Electrons
and positrons are both referred to as electrons in this paper.} beams
of energy $E_e = 27.5\gev$ and proton beams of energy $E_p = 920\ (820)\gev$, 
resulting in centre-of-mass energies of $\sqrt s=318\ (300)\gev$, 
corresponding to integrated luminosities of 430 (38)~pb$^{-1}$.

\Zdetdesc

%\Zctdmvddesc(\ZcoosysfnBeta)

In the kinematic range of the analysis, charged particles were tracked
in the central tracking detector (CTD)~\citeCTD and, for the data
taken after 2001, also in the microvertex detector
(MVD)~\citeMVD. These components operated in a magnetic field of
$1.43\Tesla$ provided by a thin superconducting solenoid. The CTD
consisted of 72~cylindrical drift chamber layers, organised in nine
superlayers covering the polar-angle\footnote{The ZEUS coordinate
system is a right-handed Cartesian system, with the $Z$ axis pointing
in the proton beam direction, referred to as the ``forward
direction'', and the $X$ axis pointing left towards the centre of
HERA.  The coordinate origin is at the nominal interaction point.  The
polar angle, $\theta$, is measured with respect to the proton beam
direction.\xspace}
region
\mbox{$15^\circ<\theta<164^\circ$}. The MVD provided polar angle coverage from $7^\circ$ 
to $150^\circ$.

\Zcaldesc

The muon system consisted of barrel, rear (B/RMUON)~\cite{brmuon} and
forward (FMUON)~\cite{zeus:1993:bluebook} tracking detectors. The
B/RMUON consisted of limited-streamer (LS) tube chambers placed behind
the BCAL (RCAL), both inside and outside the magnetised iron yoke
surrounding the CAL. The barrel and rear muon chambers covered polar
angles from 34$^{\circ}$ to 135$^{\circ}$ and from 135$^{\circ}$ to
171$^{\circ}$, respectively.  The FMUON consisted of six planes of LS
tubes and four planes of drift chambers covering the angular region
from 5$^{\circ}$ to 32$^{\circ}$.  The muon system exploited the
magnetic field of the iron yoke and, in the forward direction, of two
iron toroids magnetised to 1.6~T to provide an independent measurement
of the muon momentum.

The iron yoke surrounding the CAL was instrumented with proportional
drift chambers to form the Backing Calorimeter
(BAC)~\cite{bac:1,plucinski}. The BAC provided position information
for muon reconstruction with a resolution of few mm for two of its
coordinates and approximately 15~cm for the third, covering the full
angular range.
% The BAC provides position information
%for muon reconstruction with a resolution of approximately 15 cm in
%three dimensions and a few mm in two dimensions, in the full angular
%range.

The luminosity was measured using the Bethe-Heitler reaction
$ep\rightarrow e\gamma\:p$ with the luminosity detector which
consisted of independent lead--scintillator
calorimeter~\cite{acpp:b32:2025} and magnetic
spectrometer~\cite{nim:a565:572} systems.

%The luminosity was determined from the rate of the bremsstrahlung
%process $ep\rightarrow e\gamma\:p$~\cite{acpp:b32:2025,nim:a565:572}.

\section{Kinematics}
The four-momenta of the incoming electron and proton, and the
scattered electron and proton are denoted by $k,p,k'$ and $p'$,
respectively.  The exclusive reaction
\begin{equation}
ep \rightarrow e\Upsilon p \rightarrow e\mu^+\mu^-p
\end{equation} 
at a given $s=(k+p)^2$ for electrons and protons, is described by the
following variables:

\begin{itemize}
\item $Q^2 = -q^2 = -(k-k')^2$, the negative four-momentum squared of the exchanged 
photon;
\item $y=(q\cdot p)/(k\cdot p)$, the fraction of the  electron energy transferred to 
the hadronic final state in the rest frame of the initial-state
proton;
\item $W^2=(q+p)^2=-Q^2+2y(k\cdot p)+M_p^2$, the centre-of-mass energy squared of the 
photon-proton system, where $M_p$ is the proton mass;

\item $M_{\mu^+ \mu^-}$, the invariant mass of the $\mu^+ \mu^-$ pair.
\end{itemize}

Selected events (see Section 4) are restricted to $Q^2$ values from
the kinematic minimum, $Q^2_{\rm min}=M_e^2y^2/(1-y) \approx
10^{-9}$~GeV$^2$, for $y=0.2$, where $M_e$ is the electron mass, to a
value at which the scattered electron starts to be observed in the
CAL, $Q^2_{\rm max}\approx$ 1~GeV$^2$. The median $Q^2$ value is
$10^{-3}$ GeV$^2$ and the photon-proton centre-of-mass energy can be
expressed as

\begin{equation}
W^2 \approx 4E_p E_e y \approx 2E_p(E-p_Z),
\end{equation}

where $(E-p_Z)$ is the difference between the energy and
the longitudinal momentum of the $\mu^+\mu^-$ pair.

\section{Event selection}

Exclusive $\mu^+ \mu^-$ events in photoproduction were selected using
dedicated triggers and offline selection cuts.  At the trigger level,
at least one CTD track matched with a F/B/RMUON deposit was required.
The offline selection was~\cite{igor:2008}:

\begin{itemize}
\item CAL timing and vertex position 
consistent with a nominal $ep$ interaction;
\item two oppositely charged tracks  matched to the vertex and 
no other track in the central tracking system;
\item at least one track identified as a muon 
according to a procedure which uses information from B/RMUON, FMUON or
BAC, whenever available in a given event~\cite{bloch:2005}; if not
explicitly identified as a muon, the second track had to be
consistent with a minimum-ionising particle;
\item tracks with hits in at least 5 CTD superlayers, to ensure a good 
momentum resolution;
\item $|\eta_1 - \eta_2|\leq$~1.5, where $\eta_i$ is the pseudorapidity\footnote{
Pseudorapidity is defined as $\eta = -\ln{(\tan\frac{\theta}{2})}$.} of
a given track, to reduce the influence of the purely electromagnetic
Bethe-Heitler background;
\item transverse momentum of a track  $p_T > 1.5$~GeV;
\item  
a cut $|\pi - \theta_1 - \theta_2|\geq$~0.1, where $\theta_i$ is the
polar angle of a given track, to further reject cosmic-ray events;
\item invariant mass $M_{\mu^+\mu^-}$ in the range between 5 and 15 GeV;

\item the energy of each CAL cluster not associated to any of 
the final-state muons to be less than
$0.5$~GeV. This threshold was set to be above the
noise level of the CAL. It implicitly selected exclusive events with
an effective cut $Q^2<1$~GeV$^2$; 
\item to suppress the contamination from
proton-dissociative events, $ep \rightarrow e\Upsilon Y$, the sum of
the energy in the FCAL surrounding the beam hole had to be smaller
than $1$~GeV~\cite{igor:2008}. This corresponds to an effective cut on
the mass $M_Y$ of the dissociated system, $M_Y \lesssim$~4~GeV.

\end{itemize}

The events were selected in the kinematic range $60<W<220$~GeV.

\section{Monte Carlo simulation}

The detector and trigger acceptance and the effects due to imperfect
detector response were determined using samples of Monte Carlo (MC)
events.  Vector-meson production was simulated using the DIFFVM 2.0
generator~\cite{proc:mc:1998:396} which is based on Regge
phenomenology and the Vector Dominance Model in which the photon
fluctuates into a virtual vector meson which interacts diffractively
with the proton via Pomeron exchange. For exclusive vector-meson
production, $s$-channel helicity conservation (SCHC) was assumed. An
exponential dependence $\sim e^{bt}$ was assumed for the differential
cross section in the four-momentum-transfer squared distribution at
the proton vertex, $t$, with a slope parameter $b = 4.5$
GeV$^{-2}$, consistent with the value obtained for exclusive $J/\psi$
electroproduction~\cite{epj:c24:345,H1jpsi:2006}. The $W$ dependence
of the $\gamma\:p \rightarrow \Upsilon\:p$ cross section was
parameterised as $\propto W^\delta$, with\footnote{This value was
obtained from an iterative process of the analysis.} $\delta$ = 1.2.
Radiative corrections are of the order of 1\,\%~\cite{spiridonov-2005}
and are not included in the simulation. The DIFFVM generator
supplemented by the JETSET 7.3 package ~\cite{manual:cern-th-7112/93}
was used to estimate the value of $M_Y$ below which background events
from proton dissociation, $\gamma p \to \Upsilon Y$, do not leave
deposits in FCAL.

The background, consisting of Bethe-Heitler dimuon events, a purely
electromagnetic process, was simulated using the GRAPE v1.1k
~\cite{cpc:136:126} MC program.  All MC events were put through the
simulation of the ZEUS detector based on the GEANT
~\cite{tech:cern-dd-ee-84-1} program versions 3.13 (1996--2000) and
3.21 (2003--2007) and were analysed with the same reconstruction and
offline procedures as the data.  In addition,
corrections~\cite{bloch:2005,igor:2008} of the muon detector
efficiencies determined from a data set consisting of isolated
$J/\psi$ and Bethe-Heitler events were applied.

\section{Signal extraction}
The distributions of the $\mu^+ \mu^-$ invariant mass for the selected
events are presented in Fig.~\ref{fig1} for the two $W$ intervals,
$60<W<130$~GeV and $130<W<220$~GeV, and for the entire sample.  A
clear signal is seen around 9.5~GeV.

The mass resolution in the $\Upsilon$ resonance region is
approximately 0.2~GeV and does not allow the $\Upsilon_i$, $i$=1,2,3,
states to be resolved.  The number of signal events in each $W$
interval was determined using the following method.  The MC of the
Bethe-Heitler process was normalised to the data in a mass window not
containing resonances ($5$ to $9\gev$, $10.7$ to $15\gev$).  The
spectrum outside the resonance regions was reproduced and the
distribution was used to obtain the signal by subtracting the
background under the resonances.  It was assumed that the above three
states are produced at HERA in the proportions 0.73:0.19:0.08, as
measured in $p\bar{p}$ collisions at the Tevatron~\cite{CDFups02}.
The extracted number of signal events in the signal region 9--10.7
GeV are tabulated in Table~\ref{table1} for the different $W$
ranges. The DIFFVM generator was used to obtain the combined shape of
the three $\Upsilon_i$ states (see Fig. 1). The DIFFVM MC was
normalised to the observed number of signal events.  Using the
above-mentioned ratios, the number of $\Upsilon(1S)$ candidates was
calculated and is also presented in Table~\ref{table1}.

%This approach yielded 85$\pm$17 signal
%(62$\pm$12 $\Upsilon$(1S)) events for $60<W<220$~GeV, 41$\pm$12 signal
%(30$\pm$9 $\Upsilon$(1S)) events for $60<W<130$~GeV, and 44$\pm$13
%signal (32$\pm$9 $\Upsilon$(1S)) events for $130<W<220$~GeV
%(Table~\ref{table1}) which were later used for calculating the cross
%section.

The procedure to determine the fraction of proton-dissociative events
in the final sample, $f_{\rm pdiss}$, has been described in detail
elsewhere~\cite{epj:c24:345}. Due to the small rate of
proton-dissociative $\Upsilon$ production, it was not possible to use
the analysed data themselves to evaluate this fraction even on the
full data sample. However, this contribution is expected to be similar
in all diffractive vector meson production
processes~\cite{zeusvm:2000}. Therefore, diffractive $J/\psi$ meson
production~\cite{epj:c24:345} was used to evaluate this correction.
The value assumed in this paper was $f_{\rm pdiss}=0.25$.

The effective photon flux calculation~\cite{Flux} takes into account the
transverse and longitudinal flux factors~\cite{proc:mc:1998:396}.
Thus, the effective flux calculation depends on the assumed
parameterisation $W^{\delta}$ of  $\sigma(\gamma p)$. The
uncertainty on the $W$ dependence is taken into account as a
systematic uncertainty.

\section{Systematic uncertainties}

The following sources of systematic uncertainty were
considered~\cite{igor:2008} (numbers for the full $W$ range are
given):

\begin{itemize}

\item
signal extraction method: the background was normalised separately to
the left and to the right sides of the mass spectrum
outside of the signal region: $^{+1}_{-9}\%$;

\item the uncertainty of the CTD tracking and muon-chamber performance in the 
trigger, and the subsequent muon reconstruction in the offline
analysis: $\pm 9\%$;

\item $f_{\rm pdiss}$ was varied between 0.2 and 0.3 to account for uncertainties
in the experimental conditions and for the fact that it was evaluated
from $J/\psi$ production~\cite{epj:c24:345}, resulting in an
uncertainty of $\pm7\%$;

\item the uncertainty related to a variation of the treatment of the 
calorimeter noise: $-4 \%$;

\item the variation of $\delta = 1.2 \pm 0.5$ results in an uncertainty of 
$+3.2,-2.2\%$ on the $ep$ cross section and $ -3.9\%, +1.6\%$ on the
$\gamma p$ cross section;

\item variation of the slope parameter $b = 4.5 \pm 0.5$ GeV$^{-2}$ gives a
negligible effect on the acceptance.
\end{itemize}

In addition, the following uncertainties related to the $\Upsilon$(1S)
extraction were also considered:

\begin{itemize}
\item the $\Upsilon$(1S) fraction in the signal was varied
within the total uncertainties quoted by the CDF
collaboration~\cite{CDFups02}: $+5\%,-6\%$;

\item the $\Upsilon$(1S) decay branching ratio: $\pm 2.4\%$ ~\cite{pl:b592:1}.

\end{itemize}

The total systematic uncertainty was determined by adding the
individual contributions in quadrature. The values for the different $W$
ranges are given in Table ~\ref{table1}.

The uncertainty of the luminosity determination is 2.6 $\%$ and is not
included in the result.

\section{Results}

All $ep$ cross sections are given at $\sqrt{s}$ = 318 GeV.

For each $W$ bin, the sum of $ep$ cross sections multiplied by the
corresponding decay branching ratios to muons was evaluated according
to the formula

\begin{equation}
\sum_i \sigma^{\gamma p \rightarrow \Upsilon_i p} \cdot \mathcal{B}_i 
 = \frac{N_{\rm sig} (1 - f_{\rm pdiss})}{\cal{A}  
\cal{L}},
\end{equation}
where $N_{\rm sig}$ is the number of signal events in the signal mass
region, $\cal{A}$ is the overall acceptance, $\mathcal{B}_i$ 
is the decay branching ratio into $\mu^+ \mu^-$, and $\cal{L}$ is
the corresponding integrated luminosity.

The $ep$ cross section for $\Upsilon$(1S) production
was calculated according to 

\begin{equation}
\sigma^{ep\rightarrow \Upsilon(1S) p} = \frac{f_{\Upsilon(1S)}}{\mathcal{B}_1}
\sum_i \sigma^{\gamma p \rightarrow \Upsilon_i p} \cdot \mathcal{B}_i,
\end{equation}
where $f_{\Upsilon(1S)}$ = $0.73^{+0.05}_{-0.06}$~\cite{CDFups02} and
$\mathcal{B}_1$ = $2.48\pm 0.06 \%$\cite{pl:b592:1}.

The $\gamma p$ cross section for exclusive $\Upsilon$(1S) photoproduction was
obtained through the relation

\begin{equation}
\sigma^{\gamma p\rightarrow \Upsilon(1S) p} = \frac{1}{\Phi} \sigma^{ep\rightarrow \Upsilon(1S) p},
\end{equation}

where $\Phi$ is the effective photon flux~\cite{igor:2008}. The
following results were obtained for $Q^2 <$ 1 GeV$^2$:
\begin{eqnarray}
\sigma^{\gamma p\rightarrow \Upsilon (1S) p}& =& 160\pm 51 ^{+48}_{-21}\ \   {\rm pb}, \ \ \ \   
60 < W < 130\  {\rm GeV} \\
\sigma^{\gamma p\rightarrow \Upsilon (1S) p}& =& 321\pm 88 ^{+46}_{-114}\  {\rm pb}, \ \ 
 130 < W < 220\  {\rm GeV} \\
\sigma^{\gamma p\rightarrow \Upsilon (1S) p}& =& 235\pm 47 ^{+30}_{-40}\ \   {\rm pb}, \ \ \ \
 60 < W < 220\  {\rm GeV} 
\end{eqnarray}

The number of events, the acceptance, the flux and the cross sections
in the different $W$ intervals are given in
Table~\ref{table1}.

Figure 2 shows the extracted cross section for the two independent $W$
ranges, $60<W<130$~GeV and $130<W<220$~GeV. Also shown are a previous ZEUS
result~\cite{pl:b437:432} based on a partially overlapping data set
and the H1 result~\cite{pl:b483:23}.  The measured two cross
section values were used to calculate $\delta$, resulting in
$\delta=1.2\pm0.8$, a value consistent with the theoretical
expectation~\cite{Frankfurt:1998yf}.

The data are compared to several theoretical calculations.  Frankfurt,
McDermott and Strikman (FMS) ~\cite{Frankfurt:1998yf} based their
calculation on a two-gluon exchange between the interacting $q\bar{q}$
dipoles and the proton, using CTEQ4L parton density functions (PDFs)
~\cite{PhysRevD.55.1280}.  Ivanov, Krasnikov and Szymanowski (IKS)
~\cite{Ivanov:2004nk} use a next-to-leading-order (NLO) calculation in
which the prediction for the $W$ dependence of the cross section
depends on the scale adopted for the hard scattering (presented here
for $\mu=$1.3 and 7~GeV).  Martin, Nockles, Ryskin and Teubner (MNRT)
~\cite{Martin:2007sb} have a NLO calculation using gluon densities
extracted from HERA data on exclusive $J/\psi$ electroproduction.
Rybarska, Sch\"afer and Szczurek (RSS)~\cite{rss} use a
$k_T$-factorisation approach trying a Gaussian-like and a Coulomb-like
light-cone wave-function for the vector meson. All the calculations
are consistent with the data. In the IKS case, a calculation using an
intermediate scale is preferred. In the RSS case, the data seem to
favour a Gaussian-like wave function.

\section{Summary}

The exclusive photoproduction of $\Upsilon$(1S) meson has been studied
at HERA with the ZEUS detector in the kinematic range $60<W<220$~GeV,
$Q^2<1$~GeV$^2$ and $\sqrt s=318\gev$ using the muon decay channel.  The
dependence of the cross section on $W$ has been extracted and is in
agreement with predictions of several calculations based on
perturbative QCD.

\section*{Acknowledgments}
We appreciate the contributions to the construction and maintenance of
the ZEUS detector of many people who are not listed as authors. The
HERA machine group and the DESY computing staff are especially
acknowledged for their success in providing excellent operation of the
collider and the data-analysis environment. We thank the DESY
directorate for their strong support and encouragement.

%For each $W$ bin the $ep$ cross section for $\Upsilon$(1S) production
%was evaluated according to the formula
%
%\begin{equation}
%\sigma^{ep\rightarrow \Upsilon(1S) p} = \frac{N_{\Upsilon(1S)} (1 - f_{pdiss})}{\cal{A}  
%\cal{B} \cal{L}},
%\end{equation}
%
%where $N_{\Upsilon(1S)}$ is the number of signal events, $\cal{A}$ is
%the overall acceptance, $\cal{B}$ = $2.48\pm 0.06 \%$ is the decay
%branching ratio into $\mu^+ \mu^-$, and  $\cal{L}$ is the corresponding
%integrated luminosity.

%------------------------------------------------------------------------------
%       Bibliography
%------------------------------------------------------------------------------

%%%%%%%%%%%%%%%%%%%%%%%%%%%%%%%%%%%%%%%%%%%%%%%%%%%%%%%%%%%%%%%%%%%%%%%%%%%%%%%%%%%%%%%%%%%%%%%%%%
%%%%% Bibliography
%%%%%%%%%%%%%%%%%%%%%%%%%%%%%%%%%%%%%%%%%%%%%%%%%%%%%%%%%%%%%%%%%%%%%%%%%%%
%%%%%%%%%%%%%%%%%%%%%%%%%%%%%%%%%%%%%%%%%%%%%%%%%%%%%%%%

\providecommand{\etal}{et al.\xspace}
\providecommand{\coll}{Coll.\xspace}

%------------------------------------------------------------------------------
%       Tables
%------------------------------------------------------------------------------

%-------------------------------------------------------------------------------
%       A table with 1S
%-------------------------------------------------------------------------------

\begin{table}
\centering
\begin{tabular}{||l||c|c|c||}
\hline
\hline
$W$ range $(\rm GeV)$ & 60--130 & 130--220 & 60--220\\ \hline \hline
$N_{\rm tot}$ & $159\pm 13$ &$135\pm 12$ &$294\pm 17$\\ \hline 
$N_{\rm sig}$ & $41\pm 13$ &$44\pm 12$ &$85\pm 17$\\ \hline 
$N_{\Upsilon(\rm 1S)}$ & $ 30\pm 9 $ &$32\pm 9 $ &$ 62\pm 12 $\\ \hline 
$\cal{A}$ &0.216 & 0.230 & 0.226\\ \hline 
$\sum_i \sigma^{ e p\rightarrow \Upsilon_i p}\cdot \mathcal{B}_i \ (pb)$& 
$ 0.30\pm0.10^{+0.09}_{-0.04}$&$ 0.31\pm 0.08^{+0.04}_{-0.11}$&$ 0.60\pm 0.12^{+0.07}_{-0.09}$\\ \hline
$\sigma^{e p\rightarrow \Upsilon (1S) p} \ (pb)$& 
$8.9\pm 2.8 ^{+2.7}_{-1.2}$&$9.0\pm 2.4^{+1.3}_{-3.2}$&$17.6\pm 3.5 ^{+2.3}_{-2.9}$\\ \hline
$\Phi$ & 0.055 & 0.028 & 0.074\\ \hline
$W_0$ $(\rm GeV)$& 100 & 180 & 140\\ \hline 
$\sum_i \sigma^{ \gamma p \rightarrow \Upsilon_i p}\cdot \mathcal{B}_i \ (pb)$& 
$5.5\pm 1.8 ^{+1.6}_{-0.6} $ &$10.9\pm3.0 ^{+1.5}_{-3.8} $ &$8.1\pm 1.6 ^{+0.9}_{-1.3} $\\ \hline
$\sigma^{\gamma p\rightarrow \Upsilon (1S)  p} \ (pb)$ &
$160\pm 51 ^{+48}_{-21} $ &$321\pm 88 ^{+46}_{-114} $ &$235\pm 47 ^{+30}_{-40} $\\ \hline
\hline
\end{tabular}
\caption{The $\Upsilon$ production cross section for $Q^2<1$~GeV$^2$. 
The first uncertainty is statistical, the second systematic.
$N_{\rm tot}$ is the total number of events in the signal mass region,
$N_{\rm sig}$ is the number of signal events in the signal mass region,
$N_{\Upsilon(\rm 1S)}$ is the extracted number of $\Upsilon$(1S) signal
events, $\cal{A}$ is the acceptance, $\mathcal{B}_i$ is the decay
branching ratio to muons of $\Upsilon_i$, $\Phi$ is the effective
photon flux used to compute the $\gamma p$ cross section from the $ep$
cross section at $W_0$ and a median $Q^2_0=10^{-3} GeV^2$. }
\label{table1}
\end{table}

\setlength{\unitlength}{1mm}
%\begin{figure}[p]
%  \begin{center}
%    \begin{picture}(160,139)
%      \put(0,-3){\epsfig{file=./figs/wv.eps, width=16cm, clip}}      
%    \end{picture}
%  \end{center}
%  \caption{ $W$ distribution for the entire sample.
%  }
%  \label{fig1}
%\end{figure}

\begin{figure}[p]
  \begin{center}
    \begin{picture}(160,139)
      \put(0,-3){\epsfig{file=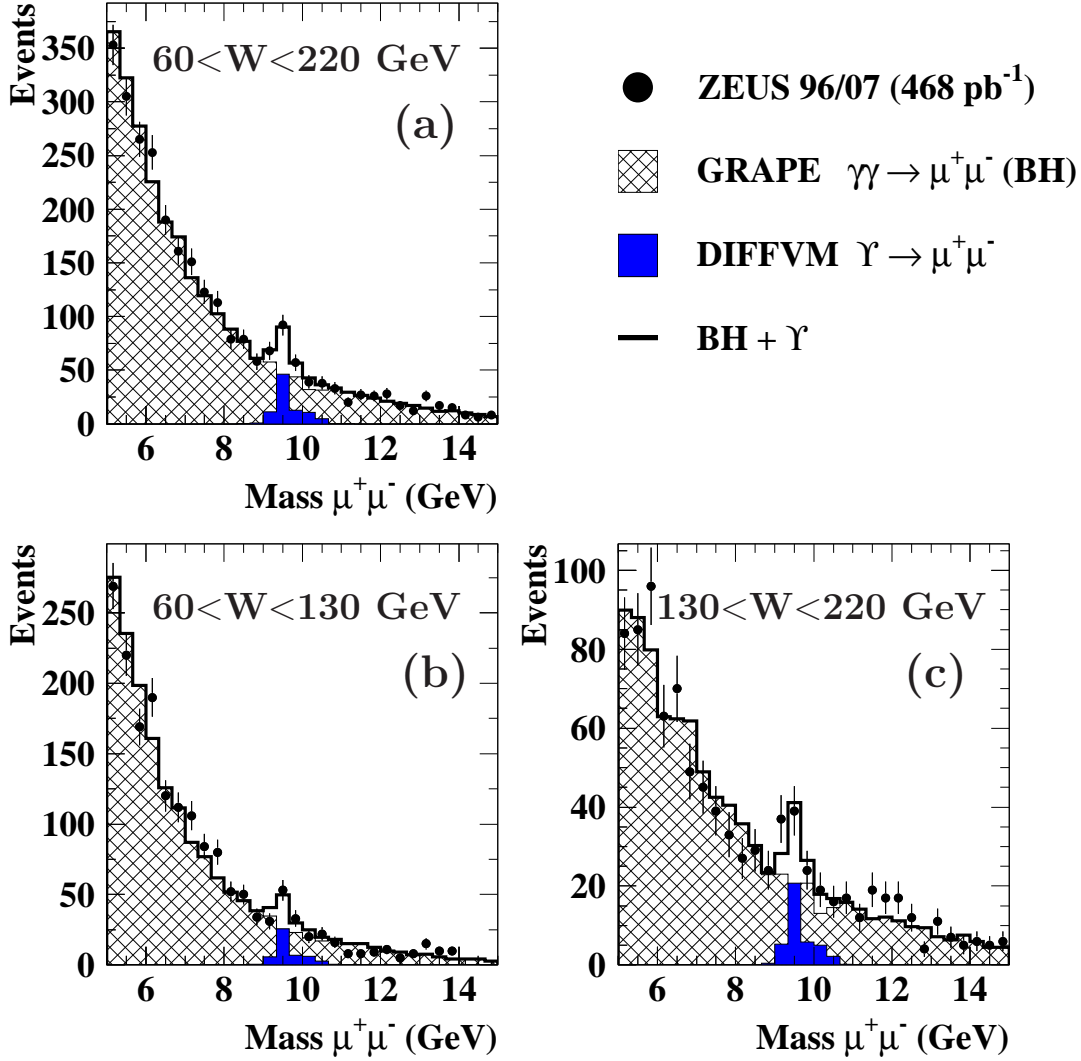, width=16cm, clip}}  
      \put(30,132){\large{\bf{60$<$W$<$220~GeV}}}
      \put(62,123){\Large{\bf{(a)}}}
      \put(30,59){\large{\bf{60$<$W$<$130~GeV}}}
      \put(63,50){\Large{\bf{(b)}}}
      \put(97,59){\large{\bf{130$<$W$<$220~GeV}}}
      \put(130,50){\Large{\bf{(c)}}}

    \end{picture} \end{center} 
\caption{ Invariant mass distributions
    of the $\mu^+ \mu^-$ pairs in different $W$ regions: $(a)$ $60<W<220$~GeV,
     $(b)$ $60<W<130$~GeV and $(c)$ $130<W<220$~GeV. The
    full dots are the ZEUS data. The hatched and full histograms
    represent, respectively: the GRAPE distribution of the
    Bethe-Heitler (BH) background and the sum of DIFFVM distributions
    of the $\Upsilon$\emph{(1S)}, $\Upsilon$\emph{(2S)} and
    $\Upsilon$\emph{(3S)} signals. The solid line is the sum of the
    two contributions (see text).}
  \label{fig1}
\end{figure}

\begin{figure}[p]
\begin{center}
    \begin{picture}(160,139)
    \put(0,-3){\epsfig{file=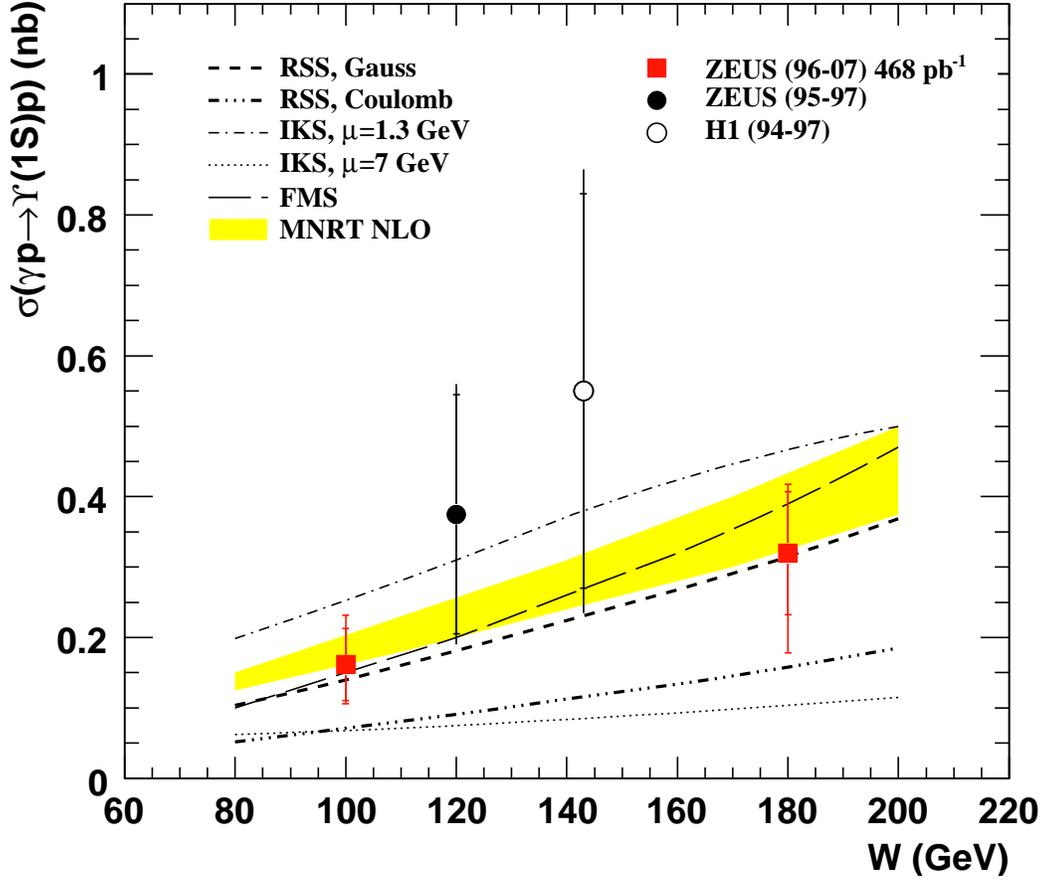, width=16cm,
    clip}} 
\end{picture} 
\end{center} 
\caption{ The exclusive
    $\Upsilon$\emph{(1S)} photoproduction cross section as a function
    of $W$. The full squares are the ZEUS data from this measurement
    in the kinematic region $Q^2<1\ GeV^2$, and two W ranges,
    $60<W<130$~GeV, and $130<W<220$~GeV.  The inner bars indicate the
    statistical uncertainties, the outer bars are the statistical and
    systematic uncertainties added in quadrature.  The earlier
    measurements of ZEUS~\emph{\protect\cite{pl:b437:432}} and
    H1~\emph{\protect\cite{pl:b483:23}}, are also shown.  The shaded
    area denotes predictions of NLO
    MNRT~\emph{\protect\cite{Martin:2007sb}}. The long-dashed line is
    the prediction of the FMS
    model~\emph{\protect\cite{Frankfurt:1998yf}}. The dashed-dotted
    (dotted) line is the prediction of the
    IKS~\emph{\protect\cite{Ivanov:2004nk}} using a scale of 1.3 GeV
    (7 GeV). The small-dashed line (small-dashed three-dots) is the
    prediction of RSS~\emph{\protect\cite{rss}}, using a Gaussian-like
    (Coulomb-like) wave function. }
\label{fig2}

~~~

~~~

~~~~

\end{figure}

%
%       ... that's it
%
\end{document}